\documentclass{aastex}
\usepackage{emulateapj5}
\begin{document}
\title{Unstable cold dark matter and the cuspy halo problem 
in dwarf galaxies}
\author{F.J. S\'{a}nchez-Salcedo }
\affil{Instituto de Astronom\'\i a, UNAM, Ciudad Universitaria, Aptdo. 70 264,
C.P. 04510, Mexico City, Mexico}
\email{jsanchez@astroscu.unam.mx}

\begin{abstract}
We speculate that the dark halos of dwarf galaxies
and low surface brightness galaxies soften their central cusps by
the decay of a fraction of cold dark matter (CDM) particles to a 
stable particle with
a recoiling velocity of a few tens km s$^{-1}$, 
after they have driven the formation of galactic halos.
This process, however, does not necessarily
produce a significant reduction of the central dark matter density
of satellite dwarf spheroidals like Draco or Fornax. 
It is shown that the recovered distribution of concentration parameters
$c$ for the initial (before decay) 
Navarro-Frenk-White halos, is in good agreement with 
CDM predictions. Other interesting potentials of unstable CDM are
highlighted.

\end{abstract}
\section{Introduction}
The hierarchical clustering cold dark matter (CDM) scenario for
structure formation has been very successful in reproducing most
properties of the local and high-redshift Universe. However, some few perceived
problems of such models, particularly at small scales, remain open.
Numerical studies of collisionless CDM with different
initial density fluctuation spectra predict
dark halos with steep cusps of the form
$\rho(r)\sim \rho_{0}(r/r_{\rm s})^{-\alpha} (1+r/r_{\rm s})^{\alpha-3}$,
with $\alpha=1$--$1.5$, after Navarro, Frenk \& White (1996) (NFW profile, 
hereafter),
whereas most of the rotation curves of dwarf
galaxies and low surface brightness (LSB) galaxies suggest that their halos
have constant density cores (see de Blok, Bosma \& McGaugh 2003,
and references therein). 
A possibly related problem is the
apparent overprediction of the number of satellites with circular
velocities $v_{\rm c}\sim 10$--$30$ km s$^{-1}$ within the virialized
dark halos of the Milky Way and M31 (Klypin et al.~1999; Moore et al.~1999).

Many astrophysical mechanisms have been suggested 
to explain these discrepancies. 
If winds driven by vigorous star formation, bars in forming
galaxies and the merger of black holes prove to be
inefficient at altering a primordial dark matter cusp (see
Gnedin \& Zhao 2002, Sellwood 2003 and Read \& Gilmore 2003, respectively),
one should invoke modifications of the standard CDM scenario.
A reduction of the small-scale power by 
substituting warm dark matter for CDM or by appealing to a model of
inflation with broken scale invariance 
is strongly discouraged by the recent 
WMAP observations (Bennett et al.~2003; Spergel et al.~2003).

Other alternatives involve 
altering the fundamental nature of the dark matter itself 
(Spergel \& Steinhardt 2000; Kaplinghat, Knox \& Turner 2000;
Cen 2001a; Giraud 2001; Arbey et al.~2003, and references therein).
All these models have free parameters tunable to fit the observations. 
For instance, the self-interacting dark matter (SIDM) scenario with a general
velocity dependent cross-section per unit mass 
$\sigma_{\rm D}/m_{\rm D}=\sigma_{0}\left(v_{100}/v\right)^{a}$,
where $v_{100}=100$ km s$^{-1}$, satisfied all the astrophysical
requirements for a very narrow range of parameters $\sigma_{0}=0.5$--$1$
cm$^{2}$ g$^{-1}$ and $a=0.5$--$1$ (Hennewai \& Ostriker 2002; 
Col\'{\i}n et al.~2002, and references therein).
This huge cross-section
has made SIDM unappealing to most WIMP and axion theorists.

There are some empirical evidence against SIDM. The central
densities of dark matter in mass models with dark matter cores 
for the elliptical galaxy NGC 4636 are higher than
expected in the SIDM scenario with
$a\leq 1$ (Loewenstein \& Mushotzky 2002). In addition,
the halo eccentricity derived for NGC 720 is in good agreement with
collisionless dark matter (Buote et al.~2002). 
Moreover, a significant reduction of the central condensation of
satellite subhalos by dark matter collisions, assisted by 
tidal stripping, would be difficult
to reconcile with the large stellar velocity dispersions observed in the
dwarf spheroidal galaxies (Stoehr et al.~2002; D'Onghia \& Burkert 2003).

The problems discussed above are also faced by other models whose
mechanisms to produce a core are based on interactions, either
elastic or inelastic, between dark matter particles. The case
of annihilating dark matter is an example.
This Letter is aimed at investigating a new variant of 
unstable dark matter, which corresponds to a modification
of the proposal by Cen (2001a),
to find out whether it is possible to overcome
the difficulties associated with previous scenarios but retaining the
desired phenomena. 

\section{Unstable Cold Dark Matter and Implications}
Although fits to the rotation curves of LSB galaxies
favor halos with cores, a fraction of them
is acceptably reproduced by divergent NFW profiles
as long as they are much less concentrated than typical halos in 
standard CDM. Cen (2001a) reconsiders the decaying dark matter 
scenario (e.g., Davis et al.~1981; 
Turner, Steigman \& Krauss 1984; 
Turner 1985; Dicus \& Teplitz 1986; Dekel \& Piran 1987; Sciama 1990),  
and suggests that the concentration of dark matter
halos lowers if one-half of the CDM particles decay to relativistic
(but nonradiative) particles by redshift $z=0$. 
The decaying CDM scenario (DCDM, hereafter)
would provide also an explanation for the formation
of dwarf spheroidal galaxies with low velocity dispersions (Cen 2001b).

Unfortunately, there are some concerns associated with the DCDM model.
First, not all the galaxies can be fitted with a NFW profile varying
the concentration parameter, especially dwarf galaxies (e.g.,
Blais-Ouellette, Amram \& Carignan 2001).
Secondly, there are empirical evidence to believe that the density of
the dark matter halo of the Milky Way at $z=2$ agrees with the present
day halo within $50$ kpc (e.g., Zhao 2002, and references therein).
The spatial distribution and dynamics of globular clusters give support
to this fact. On the other hand, a reduction in the central density
of clusters of galaxies and ellipticals by a factor $2^{4}$, as
suggested in the DCDM scenario, might be
in disagreement with observations (e.g., 
Arabadjis et al.~2002; Loewenstein \& Mushotzky 2002), 
and could spoil the good agreement found in standard CDM simulations of
halo substructure (Stoehr et al.~2002). However, we point out here that
the excessive reduction of central dark matter densities in the halos of
galaxies and clusters of galaxies can be alleviated 
if a fraction $f$ of the CDM particles decays 
to relativistic and nonradiative light particles 
plus a stable massive particle 
with a recoiling velocity in the center of mass of the
parent particle of, say $30$--$60$ km s$^{-1}$.
If the unstable parent particles are denoted by $D$ 
and the stable massive particles by $D'$, the decay is
given by $D\rightarrow D' +{\rm light}\,\, {\rm particles}$.
We will refer to this model as the unstable CDM model (UCDM).
Initially, dark halos are made up of a `primordial' 
mixture of particles $D$ and $D'$. 
Except for dark matter particles in the outer halo where the escape velocities
are smaller than the recoiling velocity, $v_{\rm rc}$, the 
particles $D'$ created by decay, will remain bound to the galactic halos.
Only the light particles that are relativistic 
will escape from the host halo.
Therefore the total mass in a Milky Way halo within $50$ kpc
would remain approximately constant (see Eq.~2-192 of Binney \&
Tremaine 1987), consistent with observations.
If $m_{D}$ and $m_{D'}$ are the particle mass of $D$ and $D'$,
respectively, a recoiling velocity of a few tens km s$^{-1}$
implies $(m_{D}-m_{D'})/m_{D}\lesssim 1$--$2\times 10^{-4}$.  
The consequences of UCDM 
will be discussed in the remainder of this Letter.

\subsection{Formation of Cores in Low-Mass Galaxies}
\label{sec:halocores}
This section is sought
to outline how the profile of dark halos would evolve when CDM particles decay
and to elucidate whether a central density core is expected.
For simplicity we will assume pure CDM halos and ignore the
contribution of the baryonic component. This is a good approximation
for galaxies dominated by the dark matter. 
The implications of the inclusion of the baryonic component
is discussed in \S \ref{sec:dSph}. 
As long as a certain dark halo formed at high redshift and the mean
lifetime of dark particles is long enough, 
the initial halo made up of a mixture of particles $D$ and $D'$,
should follow a NFW profile with an inner slope $\alpha\approx 1$--$1.5$ 
and a concentration, $c_{\rm i}$,
as predicted in cosmological simulations (Navarro et al.~1996, 1997; 
Power et al.~2003).
If $f_{D}$ is the `primordial' fraction of particles $D$ and $f_{DD'}(t)$ 
is the fraction of particles of type $D$ that have decayed at time $t$, then
the fraction of decayed particles is $f(t)=f_{D}f_{DD'}(t)$.
Note that we do not need to employ a fine-tuning between the mean lifetime
of $D$ particles, $\tau$, and the present Hubble time $t_{\rm H}$;
the case $\tau$ much shorter than $t_{\rm H}$ just implies that all
particles $D$ have already decayed at the present Hubble time, 
i.e. $f(t_{\rm H})=f_{\rm D}$.

The kinetic energy of the daughter particles $D'$ just after decay
depends on the angle between the velocity of the parent particle
and the recoiling velocity; some of them can even lose energy if, for instance, 
the recoiling velocity is opposite to the direction of the orbital 
velocity of the parent particle (except when $v_{\rm rc}\gg \sigma_{\rm t}$,
where $\sigma_{\rm t}$ denotes the typical one-dimensional
velocity dispersion of halo particles). However, the
kinetic energy of the daughter particle, just after decay, averaged over all
the possible angles of decay of a parent particle with velocity
${\mbox{\boldmath{$v$}}}_{\rm D}$ is 
$\left<E\right>=m_{\rm D'}v_{\rm D}^{2}/2+m_{\rm D'}v_{\rm rc}^{2}/2$.
Therefore, the net increase of energy is $m_{\rm D'}v_{\rm rc}^{2}/2$,
which is independent, on average,
of the initial velocity $v_{\rm D}^{2}$.

The dark matter decay will result in a gradual expansion of the halos.
The typical variation in the size of halos 
comes from virial arguments as follows. A halo with a 
virial radius $R_{\rm vir,i}$ will expand after the decay of dark 
particles to a virial radius $R_{\rm vir,f}$ of order of 
$R_{\rm vir,f}/R_{\rm vir,i}\approx (1-f x^{2}/3)^{-1}$, where
$x\equiv v_{\rm rc}/\sigma_{\rm t}$. 
For $x$ small, as occurs in clusters of galaxies (typically $x\approx
0.04$) and giant ellipticals, the effect is negligible, 
$R_{\rm vir,f}\approx R_{\rm vir,i}[1 +
{\mathcal{O}}(x^{2})]$, even if $f=1$. 
Accordingly, the profiles of massive galaxies and galaxy clusters
should not deviate significantly from the predictions of standard
collisionless dark matter in the UCDM hypothesis. However, for objects
with low $\sigma_{\rm t}$, as occurs in dwarf
galaxies, the inclusion of dark matter decay may have important 
consequences.

The detailed evolution of the density profile of a certain halo will depend
on the distribution function of its particles.
The least favorable case for core formation occurs when the
halo particles are assumed to be initially on circular orbits because
this situation minimizes the kinetic energy in the inner halo. 
By contrast, in a halo dominated by radial orbits, the circularization of
orbits by the decay could partly remove the cusp.
In order to quantify the effect of different internal kinematics
of the CDM particles, we explore first
the case in which most of the dark particles are on nearly 
circular orbits, followed by an isotropic model.

To illustrate how cores are produced, let us assume that the inner circular
velocity $v_{\rm c}(R,t)$ of the initial configuration of the halo, 
when almost no particle has decayed yet, 
is given approximately by the power law
$v_{\rm c}(R,0)=v_{0}\left(R/R_{0}\right)^{1/2\beta}$,
with $\beta> 0$ and $v_{0}$ the circular velocity at a
galactocentric radius $R_{0}$. In the NFW profile, the
exponent $1/(2\beta)\approx 1/2$, whereas for a more shallow
rotation curve, as the rigid-solid rotation, $1/(2\beta)\approx 1$. 
Suppose that between $t=0$ and
$t=dt$ a small fraction of particles decays ($f\ll 1$). 
The daughter particles `gain' an amount of energy of  
$\Delta E\approx m_{\rm D'}v_{\rm rc}^{2}/2$, as said before.
If we define the mean radius for non-circular orbits
as the radius of a circular orbit with the same energy,
it is straightforward to show that the mean radius
of the daughter particle, $R_{\rm D'}$, is:
\begin{equation}
R_{\rm D'}=\left(R_{\rm D}^{1/\beta}+\frac{1}{2\beta+1}
\left(\frac{v_{\rm rc}}{v_{0}}\right)^{2} R_{0}^{1/\beta}\right)^{\beta},
\end{equation}
where $R_{\rm D}$ is the mean radius of its parent particle.
A {\it measure} of the expansion of the halo, when the initial
orbits are approximately circular, is given by
the relative variation in the radius of the particles:
\begin{equation}
\frac{R_{\rm D'}-R_{\rm D}}{R_{\rm D}}=\sum_{i=1}^{\infty}
\pmatrix{\beta \cr i \cr}\left(\frac{1}{2\beta+1}\right)^{i} 
\left(\frac{v_{\rm rc}}{v_0}\right)^{2i}
\left(\frac{R_{0}}{R_{\rm D}}\right)^{i/\beta},
\label{eq:ratio}
\end{equation}
which goes as $\propto R_{\rm D}^{-1/\beta}$ 
in the particular case that $v_{\rm rc}\lesssim v_{0}$ 
and $R_{\rm D}\gtrsim R_{0}$,
keeping only the leading term. 
This result suggests that the rotation curve becomes more shallow,
once the fraction of mass that decays is not longer negligible at
say $t=\delta t$
\footnote{In the DCDM scenario proposed by Cen (2001a),
the byproducts of the decay were assumed to be relativistic. 
This assumption immediately implies that the adiabatic evolution
of the halo is self-similar and hence no core is formed. By contrast,
the inclusion of a recoiling velocity breaks down the self-similarity
in the evolution.}.
Simultaneously, the halo will undergo an
adiabatic expansion as a response to the new mass configuration,
decreasing the velocity dispersion of the halo particles.
This expansion also contributes to the formation of a shallow profile.
Inserting the new values of $\beta$ and $v_{0}$ at
$t=\delta t$ in Eq.~(\ref{eq:ratio}), 
we see that $1/(2\beta)$ always increases in time, concluding
that the formation of a core should be a natural consequence in this model.
For a more realistic, isotropic orbital distribution, the formation of
a core should be even more favorable.

The procedure outlined above allows us to calculate numerically
the evolution of a dark halo starting with a NFW profile 
($\alpha=1$) and for different values of 
$v_{\rm rc}$ (see Fig.~1). For the isotropic models, we have taken
advantage of the analytical fit to the distribution function given
by Widrow (2002). 
The initial halo density profile
is defined by the concentration parameter $c$
and the rotation velocity $V_{200}$ at radius $R_{200}$, that is 
the radius inside which the average overdensity is $200$ times the
critical density of the Universe. The resulting rotation curves should
be regarded as lower limits since the orbits  
of the daughter particles were assumed to be circular. However, the 
maximum increment of the rotation velocity by including an isotropic
orbital distribution for these particles can be estimated from 
the differential energy distribution (e.g., Binney \& Tremaine 1987) 
and it is marked with error bars in Fig.~1.

In order to quantify the relevance of the UCDM model, we 
have recovered the concentration parameter before CDM decay, $c_{i}$,
for three extreme galaxies: IC 2574, NGC 3274 and Draco.
IC 2574 is the prototype dwarf LSB with a conspicuous core 
(e.g., Blais-Ouellette et al.~2002) 
whereas NGC 3274 is a LSB galaxy with one of the smallest core-radius,
which can be fitted with a NFW profile (de Blok \& Bosma 2002). 
On the other hand, Draco is a dwarf spheroidal (dSph) and presents
central dark matter densities (within a radius of $\sim 1$ kpc)
as high as those predicted by standard
collisionless CDM models (Kleyna et al.~2001; 
Lokas 2002; Stoehr et al.~2002; Hayashi et al.~2003).
Therefore, these galaxies should bracket the concentration 
parameters for low-mass galaxies.
If the initial concentration parameters for these extreme galaxies, $c_{i}$,
lie within 1$\sigma$ scatter of the predictions of
standard CDM simulations, 
it would imply that our model is fully consistent with the
observed distribution of concentrations.
For the reference values $f=0.5$ and $v_{\rm rc}=40$ km s$^{-1}$, 
the observed rotation curves are reproduced well in the isotropic model, 
starting with a NFW halo with
$c_{i} \approx 13, 26, 20$ and $V_{200,i}\approx 60, 55, 60$ km s$^{-1}$ 
for IC 2574, NGC 3274 and Draco, respectively. 
Although cosmological N-body simulations must be done for a fully
description of the profile of UCDM halos,
these results are promising if one considers that expected values of
$c_{\rm i}$ in $\Lambda$CDM cosmology have $\sigma(\log c_{\rm i})=0.18$ and
a $2\sigma$ range from $5$ to $25$, with an average of around
$10$ to $15$ (NFW; Eke et al.~2001). Since the effects of UCDM are smaller
for more massive halos, the present mechanism for core formation is fully 
consistent with the fact that
the best-fitting models of high surface brightness galaxies favor
cores with small sizes (Jim\'{e}nez, Verde \& Oh 2003).

\subsection{The Case of Dwarf Spheroidals}
\label{sec:dSph}
Baryonic effects on the dark matter distribution could
affect our estimate of $c_{i}$. In particular,
if the progenitors of dSph were much more massive than presently, 
the enhancement of the velocity
dispersion of CDM particles by adiabatic compression would have effectively
reduced the evacuation of dark matter particles.
In the following we will analyze how massive the dSph progenitors
should have been to {\it suppress} significantly the effect of CDM decay.
For simplicity, the baryons are assumed to dominate the total
mass (dark matter plus baryons) within the optical radius of the
galaxy during the phase of CDM decay. The corresponding density
profile in this phase is taken to be $\rho\sim R^{-\xi}$, with
$\xi\approx 1.8$--$2$, for consistency with other
studies (e.g., Hennawi \& Ostriker 2002). The flat circular velocity
associated with this configuration will be denoted by $v_{\rm c}$. 
Under the approximation that the CDM particles are on circular orbits, 
a CDM particle initially
at a orbital radius $R$ will shrink to a radius $R'=(M/M')R$, where
$M$ is the initial mass within $R$ and $M'$ is the final mass
within $R'$, during the dissipative collapse of baryons. 
After decay, the daughter particle will acquire a mean circular radius
$R(M/M')\exp\left(v_{\rm rc}^{2}/2v_{\rm c}^{2}\right)$.
Later on, most of the baryonic mass will be ejected probably
by supernova explosions or stripped by ram pressure. 
This mass loss will expand the orbit of the daughter particle to a final
radius of $\sim R\exp\left(v_{\rm rc}^{2}/2v_{\rm c}^{2}\right)$,
still assuming that the process evolves adiabatically. The net
effect is a self-similar expansion $R\rightarrow \lambda R$, with
$\lambda\equiv \exp\left(v_{\rm rc}^{2}/2v_{\rm c}^{2}\right)$ and,
therefore, the initial inner velocity profile $v(R)\propto R^{1/2}$,
associated with a NFW profile $\rho_{\rm D}\propto R^{-1}$, is 
transformed into $v'(R)=\nu v(R)$ with $\nu^{2}=1-f+f/\lambda^{2}$. 
We will employ a variation of $\nu\approx 0.8$, which 
is compatible with observational uncertainties ($> 25 \%$)
in the mass determination of dSph (e.g., Kleyna et al.~2001; Lokas 2002), 
and the fact that, in our model, the mass loss of CDM particles induced
by Galactic tidal forces would be significantly suppressed
if the progenitor was more compact in the past. For $f=0.5$, 
$v_{\rm rc}=40$ km s$^{-1}$ and $\nu=0.8$, a value 
$v_{\rm c}\gtrsim 35$ km s$^{-1}$ is required. 
In terms of the baryonic mass,
this implies a mass of gas $\gtrsim 2.5\times 10^{8}$ M$_{\odot}$ 
for the progenitor of the dSph Fornax with a typical radius of $\sim 0.8$ kpc, 
and $\gtrsim 0.4\times 10^{8}$ M$_{\odot}$ for the 
progenitor of Draco with radial size of $\sim 0.15$ kpc. 
As we have assumed circular orbits, these estimates are clearly
lower limits.  For progenitors with these masses, the decay of 
UCDM could not have observable effects in their halos.
The inferred masses are large, especially for Draco,
but are not unreasonable; they are in accordance with
different lines of research that point out masses of $10^{9}$--$10^{10}$
M$_{\odot}$ for the progenitors of those dwarf spheroidals
with a long period of star formation like Draco and Fornax
(Kitayama et al.~2001; Ikuta \& Arimoto 2002). 
In a much more speculative scenario, 
if the gas mass of dSph is never larger than $10^{7}$ M$_{\odot}$, 
dSph could preserve their initial
dark halos intact under the assumption that CDM only decays
by interactions with baryons/nucleons with a certain cross-section 
$\sigma_{\rm DN}$.

\section{Discussion}
The possibility that dark matter decays is being a topic of active work.
Theoretical predictions for lifetime of heavy sterile neutrinos
and Ultra Heavy particles,
as WIMPZILLAs, cover a large range of values, from $10^{-2}$
to $10^{11}$ Gyr. We have assumed that UCDM and its byproducts
are weakly interacting particles. The possibility that
relativistic light particles of the decay are the
source of Ultra High Energy Cosmic Rays or responsible
for the reionization of the Universe is still open.
Even if Ultra Heavy particles were to decay to Standard Model
particles, they are not ruled out by present observations
(e.g., Ziaeepour 2000).

Besides the formation of density cores, other interesting
phenomena associated with UCDM arise. 
UCDM enhances substructure destruction within galactic
halos, and the evacuation of part of the central mass of dark matter in
dwarf ellipticals and dwarf spheroidals
would lead to an adiabatic expansion of their globular cluster systems,
and would partly compensate the drag of dynamical friction,
alleviating the problem of the survival of the globular clusters
(Lotz et al.~2001, and references therein). 
Moreover, assuming that UCDM particles are self-interacting
with the cross-section found by Hennawi \& Ostriker (2002),
$\sigma_{DD}/m_{D}\approx 0.02$ cm$^{2}$ g$^{-1}$,
it is feasible to incorporate the growth of supermassive
black holes in our model, as suggested by Ostriker (2000). 
The decay of particles can contribute to feed up the central black
hole if they are scattered into the loss-cone, but this contribution
is much smaller than the Bondi accretion.

\acknowledgements
I thank V.~Avila-Reese and an anonymous referee for helpful comments.

\clearpage
\begin{figure}
\plotone{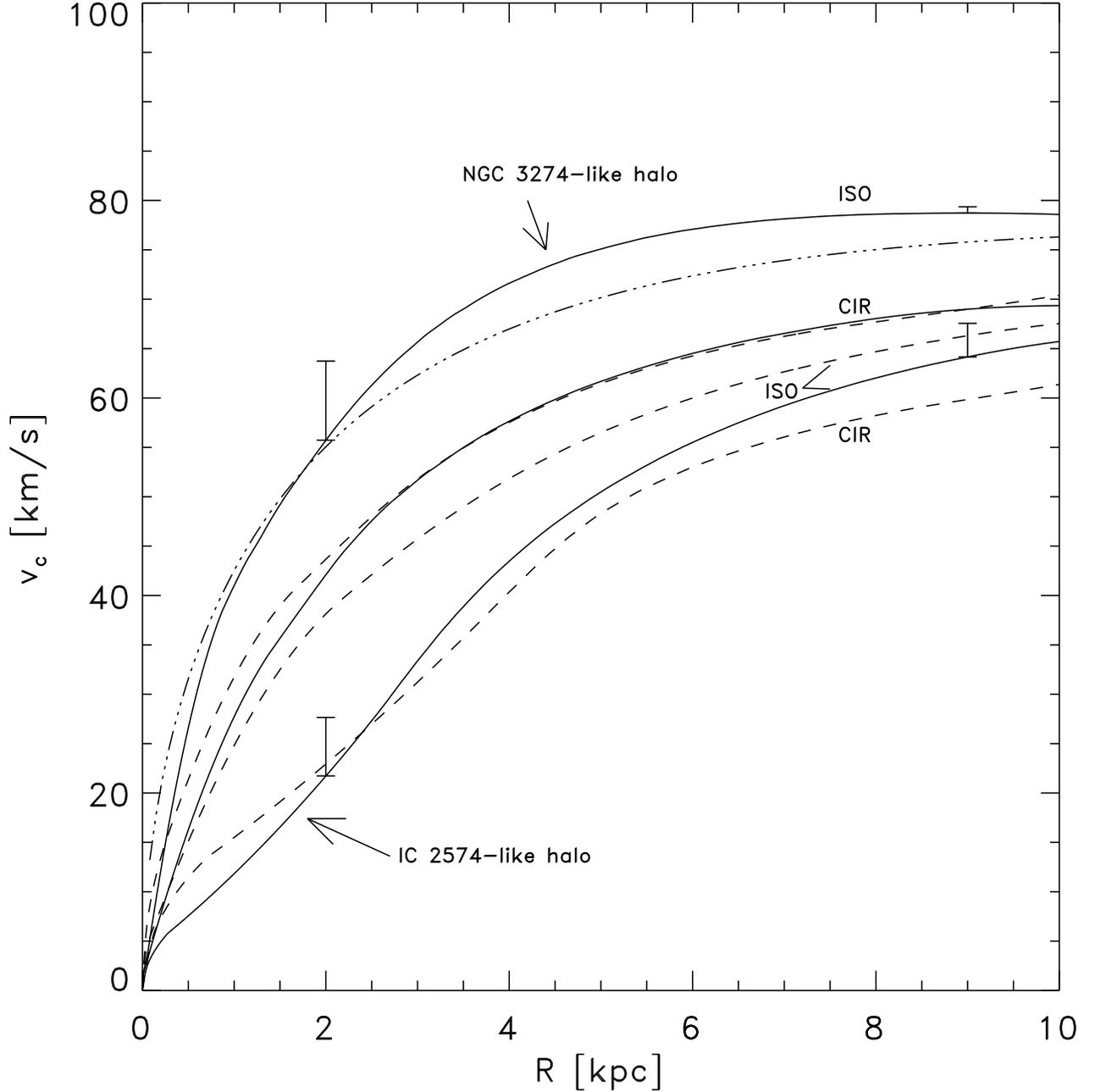}
\caption{\footnotesize Variations in the shape of the halo
contribution to the rotation curve 
for different values of $f$ and $v_{\rm rc}$, starting with a
NFW profile with $c_{\rm i}=13$ and $V_{200,i}=60$ km s$^{-1}$ (triple-dotted
dashed line) except for the NGC 3274 halo for which $c_{\rm i}=26$
and $V_{200,i}=55$ km s$^{-1}$. Isotropic models are labeled by ISO,
while models in which all particles are initially on circular orbits
are indicated by CIR. The three solid lines show the rotation
curves for $f=0.5$ and $v_{\rm rc}=40$ km s$^{-1}$. The dashed lines
correspond to $v_{\rm rc}=60$ km s$^{-1}$ and $f=0.5$ (lower dashed
curve) and $f=0.25$ (the two upper dashed curves). The bars account
for the maximum effect when including non-circular orbits for
the daughter particles for the IC 2574 and NGC 3274 halos.}
\label{fig:fig1}
\end{figure}
\end{document}